%% file: LGTinQEDv2.tex
\documentclass[12pt]{article}

\pdfoutput=1
\usepackage[colorlinks=true,linkcolor=darkblue,citecolor=darkblue,urlcolor=darkblue]{hyperref}
\usepackage[T1]{fontenc}
\usepackage[ansinew]{inputenc}
\usepackage[english]{babel}
\usepackage{color}
\usepackage{epsfig, palatino}
\usepackage{pstricks,pst-node,pst-tree}
\usepackage{epic}
\usepackage{mathrsfs}
\usepackage{ae} 
\usepackage{amsmath}
\usepackage{amssymb}
\usepackage{graphicx}
\usepackage{ulem}
\usepackage{color}
\definecolor{darkblue}{cmyk}{0.9,0.9,0,0}
\usepackage{cite}
\usepackage{wasysym}
\usepackage{varioref}
\usepackage{makeidx}
\usepackage{simplewick}
\usepackage{array}
\usepackage{mathrsfs}
\usepackage{feynmf} 
\usepackage{tikz}
\usepackage{dsfont}
\usepackage{slashed}
\usepackage{subfig}

\usetikzlibrary{decorations.pathmorphing}

\newcommand{\comment}[1]{}

\newcommand{\beq}{\begin{equation}}
\newcommand{\eeq}{\end{equation}}
\newcommand{\beqq}{\begin{equation*}}
\newcommand{\eeqq}{\end{equation*}}
\newcommand\beqa{\begin{eqnarray}}
\newcommand\eeqa{\end{eqnarray}}
\newcommand\beqaa{\begin{eqnarray*}}
\newcommand\eeqaa{\end{eqnarray*}}
\newcommand\bea{\begin{array}}
\newcommand\eea{\end{array}}

\newcommand{\nn}{\nonumber}

\newcommand{\neqa}{\nonumber\end{eqnarray}} 
\newcommand{\la}[1]{\label{#1}}

\newcommand{\rf}[1]{(\ref{#1})}

\renewcommand{\d}{\partial}

\newcommand{\<}{{\langle}}
\renewcommand{\>}{{\rangle}}

\newcommand{\cA}{{\cal A}}

\newcommand{\re}{\relax{\rm I\kern-.18em R}}

\renewcommand{\sp}{p\hspace{-.40em}/}

\newcommand{\Blue}[1]{{\color{blue}#1\color{blue}}}

\def\XXint#1#2#3{{\setbox0=\hbox{$#1{#2#3}{\int}$}
\vcenter{\hbox{$#2#3$}}\kern-.5\wd0}}

\def\[{\left[}
\def\]{\right]}

\def\e{\epsilon}
\def\ve{\varepsilon}

\def\({\left(}
\def\){\right)}
\def\[{\left[}
\def\]{\right]}

\def\<{\langle}
\def\>{\rangle}

\def\i2{\frac{i}{2}}

\def\spi{\relax{\rm \pi\kern-0.5em /}}
\def\sA{\relax{\rm A\kern-0.5em /}}
\def\sp{\relax{\rm p\kern-0.5em /}}
\def\sd{\relax{\rm \d\kern-0.5em /}}
\def\sk{\relax{\rm k\kern-0.5em /}}
\def\sn{\relax{\rm n\kern-0.5em /}}
\def\sl{\relax{\rm l\kern-0.5em /}}
\def\sP{\relax{\rm P\kern-0.7em /}}
\def\sBethe{\relax{\rm \Bethe\kern-0.5em /}}

\def\2F1{\,_2{\rm F}_1}

\def \gam {{\gamma_{z\bar z}}}
\def \gamk {\gamma_{z_k\bar z_k}}

\def \zbz {\(z, \bar z\)}

\def \Ascr {\cA^{\,^{\kern-0.2em (\infty)}}}

\makeindex

\numberwithin{figure}{section}

\begin{document}

\thispagestyle{empty}

\setcounter{page}{1}
\setcounter{footnote}{0}
\setcounter{figure}{0}
\begin{center}
$$$$
{\Large\textbf{\mathversion{bold}
		Large Gauge Symmetries and Asymptotic States in QED}\par}

\vspace{1.0cm}

\textrm{Barak Gabai and Amit Sever}
\\ \vspace{1.2cm}
\textit{School of Physics and Astronomy, Tel Aviv University, Ramat Aviv 69978, Israel}  
\vspace{4mm}

\par\vspace{1.5cm}

\textbf{Abstract}\vspace{2mm}
\end{center}
Large Gauge Transformations (LGT) are gauge transformations that do not vanish at infinity. Instead, they asymptotically approach arbitrary functions on the conformal sphere at infinity. Recently, it was argued that the LGT should be treated as an infinite set of global symmetries which are spontaneously broken by the vacuum. It was established that in QED, the Ward identities of their induced symmetries are equivalent to the Soft Photon Theorem. In this paper we study the implications of LGT on the S-matrix between physical asymptotic states in massive QED. In appose to the naively free scattering states, physical asymptotic states incorporate the long range electric field between asymptotic charged particles and were already constructed in 1970 by Kulish and Faddeev. We find that the LGT charge is independent of the particles' momenta and may be associated to the vacuum. The soft theorem's manifestation as a Ward identity turns out to be an outcome of not working with the physical asymptotic states.

\noindent

\newpage

\tableofcontents

\parskip 5pt plus 1pt   \jot = 1.5ex

\section{Introduction}
{Recently, a new interpretation of Large Gauge Transformations (LGT) as global symmetries of QED was put forward in a series of publications \cite{mass-lessQED,massQED,othermassQED,Y-M,Y-M-old,grav_1,grav_2,Balachandran:2014hra}. LGT symmetries are generated by the $U\(1\)$ gauge transformations that asymptotically approach an arbitrary function on the conformal sphere at future (past) null infinity. The symmetry is spontaneously broken by the vacuum. Its physical implication for the S-matrix between bare asymptotic states was shown to be equivalent to the Soft Photon Theorem.\footnote{Soft theorems relate (linearly) the divergence of any scattering amplitude that has an external soft particle to the scattering amplitude without that soft particle.} Moreover, the idea that a new conservation law admitted by such a global symmetry may have implication for the entropy of black holes in GR was studied in \cite{BH-paradox}.} 

Most of the study on the subject so far focused on the S-matrix between bare (undressed) asymptotic states. However, scattering amplitudes in perturbative QED and gravity suffer from infrared divergences. These well understood divergences arise from our inability to distinguish between a ``hard'' particle with or without ``soft hair''.\footnote{In non-abelian theories there are also collinear IR divergences. As these are absent in QED (and also in gravity) we will not discuss them here.} Hence, a physical asymptotic state entering any scattering process is a super-position of the hard particle with all possible soft hair arrangements. The way to construct the physical asymptotic states that yield an IR safe S-matrix was put forward by Kulish and Faddeev \cite{kulish}. In their construction, the dressed incoming and outgoing states diagonalize the interacting asymptotic Hamiltonian. Once the S-matrix of these dressed electrons is expanded in perturbation theory, the cancellation of IR divergences is ensured by the existence of soft theorems.

The purpose of the current paper is to investigate the action of LGT on physical, i.e. dressed, asymptotic states. We find that the LGT charge can be arbitrarily distributed between the particles and the vacuum. In particular, it depends only on the particles' electric charge and is independent of their momenta. To leading order in the soft limit, the amplitude is not singular and the soft theorem becomes trivial. We leave the investigation of sub-leading orders to future work.

The paper is organized as follows. In section \ref{sec_deflgt} we start with a short review of the LGT, continue to calculate the transformation properties of asymptotic ``undressed'' states, and then, ``dress'' the asymptotic states. In section \ref{sec_lgtvac}, we study the LGT of the vacuum inside a physical scattering process. Finally, in section \ref{sec_dressed} we calculate the S-matrix between an incoming state and the LGT of an outgoing state. We show that the LGT of the dressed particles only depends on their electric charge. Specifically, it is independent of their momenta. We end this section with a Wilson line intuitive picture explaining this effect.

While we were at the final stage of writing this paper, e-print \cite{other} authored by Mirbabayia and Porrati appeared which partly overlaps with the work presented here.
\section{Review} \la{sec_deflgt}
In this section we review the essential background material and use it to establish our notations. We start with a definition of LGT in QED and then work out their action on a bare (undressed) massive charged scalar particle.\footnote{Though for simplification we consider a scalar particle, our method, and therefore results, translate trivially to a spin half electron.} A reader familiar with the topic of LGT can move directly to section \ref{sec_lgtvac}.

We will mostly work in the retarded coordinates parametrization of flat four dimensional space-time. These are convenient when discussing the isometries of future null infinity, where future asymptotic photons are localized (see figure \ref{Penrose_Null}). The transformation between these and the Cartesian coordinates is,
\beq\la{retarded}
r^2=x_1^2+x_2^2+x_3^2\ ,\qquad u=t-r\ ,\qquad z = {x_1+ix_2 \over r+x_3}\\
\eeq
while the inverse coordinate transformation reads,
\beq\la{inversetrans}
t=u+r\ ,\quad  {\bf x}=r\,\hat {\mathbf x}_{z,\bar z}\equiv {r\over 1+z\bar z}\(z+\bar z, i\(\bar z-z\), 1-z\bar z\)
\eeq
here, and throughout this paper, we use bold face letters to represent three-vectors. In these coordinates the flat spacetime metric, $ds^2=-dt^2+dx_1^2+dx_2^2+dx_3^2$, takes the form  
\beq
ds^2=-du^2 - 2du\,dr + 2r^2\gamma_{z\bar z}\,dz\, d\bar z\ ,\qquad\text{where}\qquad\gamma_{z\bar z}={2 \over \(1+z\bar z\)^2}
\eeq
is the metric on the projective plane. 

\begin{figure}[t]
\centering
\def\svgwidth{12cm}
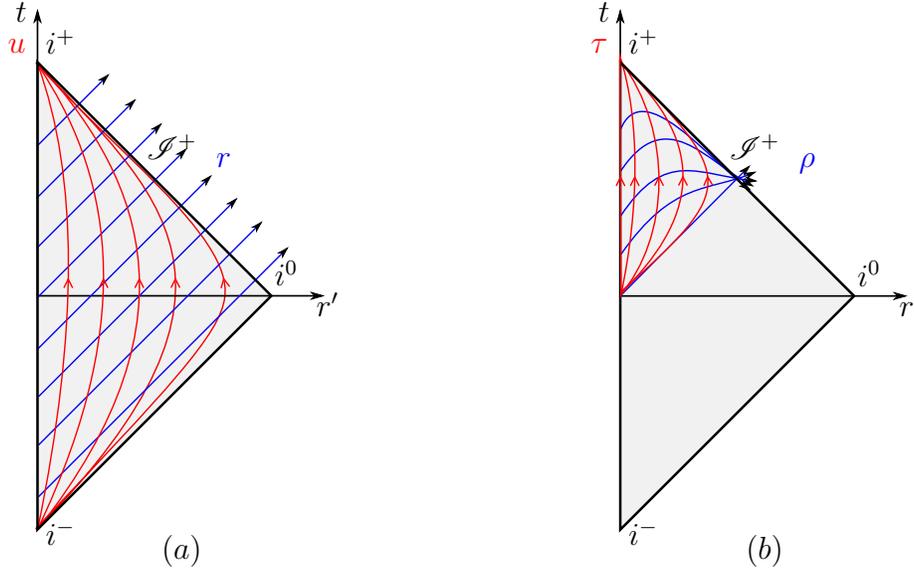
\caption{a) A Penrose diagram of flat space where $r'$ and $t$ are the radial and time coordinates. In blue are lines of constant retarded time $u$, and in red are lines of constant retarded $r$. b) The Rescaled Radial coordinates $(\rho,\tau)$ cover the patch $t>r$. In blue are lines of constant $\tau$ and in red are lines of constant $\rho$.}\la{Penrose_Null}
\end{figure}
Next, we discuss Large Gauge Transformations (LGT). 
These are the residual gauge transformations allowed by the gauge choice. When working in Lorentz gauge, $\nabla^\mu \mathcal A_\mu=0$, LGT are parametrized by a function satisfying the appropriate wave equation. Under such transformations,
\beq \la{cond-lambda}
\mathcal{A}_\mu \to \mathcal{A}_\mu + \d_\mu \lambda\ , \qquad \text{where}\qquad
\d_\mu \d^\mu \lambda = 0\ .
\eeq
If one demands that $\mathcal{A}_\mu$ will vanish at spatial infinity then $\lambda = 0$ is the only solution.  LGT come about when we relax this condition (hence the name ``Large''). More precisely, we impose the following (looser) boundary conditions at null infinity\footnote{Note that $r\to\infty$ at fixed $u$ is future null infinity, see figure \ref{Penrose_Null}.}
\beqa \la{fall-offs-a}
\lim_{r \to \infty}\mathcal{A}_r = O\(r^{-2}\)\ ,\qquad
\lim_{r \to \infty}\mathcal{A}_u = O\(r^{-1}\)\ ,\qquad
\lim_{r \to \infty}\mathcal{A}_z = O\(1\)
\eeqa
and similarly at past null infinity. This enables LGT that approach an arbitrary function on the conformal sphere,\footnote{Evidently, these are the same LGT as discussed in \cite{massQED,mass-lessQED}, where Radial gauge was chosen. The reason we chose to work in a different gauge is that in Radial gauge the LGT do not have smooth continuation to all spacetime. This makes the procedure utilized in section \ref{massive-trans}, which defines the transformation properties of massive electrons, problematic (see \cite{othermassQED}).}
\beq \la{trans-null}
\lim_{r \to \infty}\lambda\(u,r,z,\bar z\) = \ve\zbz + O\({r^{-1}}\)
\eeq 
Under such transformations, the asymptotic value of the gauge field transforms as
\beq \la{cond-lambda-null}
\Ascr_{z/\bar z} \to \Ascr_{z/\bar z} + \d_{z/\bar z} \ve\zbz\ ,\qquad\text{where}\qquad\lim_{r \to \infty} \mathcal{A}_{z/\bar z} \equiv \Ascr_{z/\bar z}+O\(r^{-1}\)\ .
\eeq
Correspondingly, the generator of the LGT at future null infinity $Q_\ve^+$ acts as
\beq \la{commute}
\[\Ascr_{z/\bar z}\(u,z,\bar z\), Q_\ve^+\] = -i\partial_{z/\bar z} \,\ve\zbz
\eeq
Similar results can be obtained for the LGT at past null infinity. There is a one to one antipodal map between the LGT of past and future null infinity. Thus, there are no additional independent transformations in past null infinity. Although in this paper we treat future null infinity exclusively, all of our results can be derived for past null infinity in the same exact manner.\footnote{This choice has the advantage that particles propagate towards the specified direction at future null infinity. As a result, some of the equations are more aesthetic. It has the disadvantage of working with Bra states instead of Ket states.}

\subsection{LGT of a Photon}
An asymptotic outgoing photon must reach future null infinity. Explicitly, the annihilation operator of a photon can be written as (see Appendix \ref{AsymFieldAppe} {for a derivation and a similar relation for the creation operator})
\beq\la{Phot-main}
a_{+/-}\(\mathbf k\) = {4\pi i \over \sqrt{\gamk}}\int du\, e^{i\omega_k u} \Ascr_{z/\bar z}\(u, z_k, \bar z_k\)
\eeq
where $(+/-)$ are the two physical polarizations of the photon and $z_k=(k_1+ik_2)/(|{\bf k}|+k_3)$. These can be chosen to be (expressed here in the $\(t,x_1,x_2,x_3\)$ coordinates)
\beq\la{polarizations}
\e_-^{\mu}\(\mathbf k\)={1 \over \sqrt 2}\(z_k,1,i, - z_k\)\qquad\text{and}\qquad\e_+^{\mu}\(\mathbf k\)={1 \over \sqrt 2}\(\bar z_k,1,-i, -\bar z_k\)\ .
\eeq
This choice is the source of correlation between the polarization vectors and the $z, \bar z$ components of the gauge field. Using (\ref{commute}) and (\ref{Phot2}) we read the LGT of the outgoing photons
\beq \la{commutation_photon_0}
\[a_+(\mathbf k),Q_\ve^+\] = {8\pi^2 \over \sqrt {\gamma_{z_k\bar z_k}}}\delta\(\omega_k\) \partial_{z} \ve\(z_k,\bar z_k\)
\eeq

\subsection{LGT of an Undressed Particle} \la{massive-trans}
Next, we study the transformation of an undressed charged massive particle. For the sake of simplicity, we take this particle to be a scalar (it is apparent that a spin half electron will transform in the same way). Unlike photons, massive particles do not reach null infinity. Thus, it will be convenient to adopt a more suitable coordinate set. We consider the so called ``Rescaled Radial Coordinates'' (RRC) (see \cite{othermassQED} and figure \ref{Penrose_Null}) 
\beq
\tau \equiv \sqrt{t^2 - r^2}=\sqrt{u^2 +2ur}\ ,\qquad\rho \equiv {r \over \sqrt{t^2 - r^2}}={r \over \sqrt{u^2 +2ur}}\ .
\eeq
The inverse coordinate transformation reads  
\beq \la{inv_RRC}
u = \tau\sqrt{1+\rho^2}-{\rho\, \tau}\ ,\qquad r = {\rho\, \tau}\ .
\eeq
In terms of the Rescaled Radial Coordinates $\tau$ and $\rho$, the flat space-time metric takes the form
\beq
ds^2 = -d\tau^2 + \tau^2\({d\rho^2 \over 1+\rho^2}+\rho^2\gam\, dz\,d\bar z\)\ .
\eeq
These coordinates are engineered so that the large $\tau$ limit with $\rho$ and $\hat {\mathbf x}$ fixed coincides with the {trajectory} of a massive particle approaching future time-like infinity. It's worth mentioning that this coordinate system will fit particles of any mass, since the trajectory is determined by $|\mathbf p| / m$. In particular, these coordinates allow us to represent the state of the outgoing massive particles on a late constant $\tau$ slice.

In this coordinate system, a Large Gauge Transformation is parametrized by a function $\lambda(\tau,\rho,z,\bar z)$ satisfying our Lorentz gauge condition 
\beq\la{cond-tau}
\d_\mu \d^\mu \lambda=\({\Delta_{\(\rho\)} \over \tau^2}-\d^2_\tau\)\lambda = 0\ ,
\eeq
where
\beq
\Delta_{\(\rho\)} \equiv \(1+\rho^2\)\d^2_\rho + {1 \over \rho}\(2+3\rho^2\)\d_\rho+{1 \over \rho^2}\(1+z\bar z\)^2\d_z\d_{\bar z}\ .
\eeq
As is evident from (\ref{inv_RRC}), the $\rho\to\infty$ boundary of any constant $\tau$ slice is the same $u=0$ point at future null infinity. Therefore, at large $\rho$ the LGT function $\lambda$ has to agree with $\ve\zbz$, which is independent of $\tau$. To work out the LGT of a massive particle we need to consider also the continuation of $\varepsilon(z,\bar z)$ to finite $\rho$ at late times
\beq
\lim_{\tau \to \infty}\lambda\(\tau, \rho,z,\bar z\)\equiv \tilde \ve\(\rho,z,\bar z\) + O\(\tau^{-\eta}\)\ ,
\eeq
where $\eta>0$ and $\lim\limits_{\rho \to \infty}\tilde \ve\(\rho,z,\bar z\)=\ve\(z,\bar z\)$. It follow from (\ref{cond-tau}) that
\beq
\Delta_{\(\rho\)}\tilde \ve\(\rho,z,\bar z\) = 0
\eeq
The solution for these equations has been worked out in \cite{othermassQED}. It is given by an integral over the conformal sphere at null-infinity of the Lienard-Wiechert electric field that is emitted by a particle with relativistic speed $\mathbf \beta = {\rho \over \sqrt{1+\rho^2}}\hat {\mathbf x}_{z,\bar z}$ and sensed approaching time-like infinity. Explicitly, we have
\beq
\tilde \ve\(\rho,z,\bar z\)= \int {d^2w\over4\pi} {{\gamma_{w,\bar w}}\,\ve(w,\bar w)\over \(\rho\,\hat {\mathbf x}_{z,\bar z} \cdot \hat {\mathbf x}_{w,\bar w}-\sqrt{1+\rho^2}\)^2}\ ,
\eeq
where $\hat {\mathbf x}_{z,\bar z}$ is the unit three-vector parametrized by $z$ as in (\ref{inversetrans}) and the dot stands for scalar product.

Now we may consider a canonically quantized (charged) scalar field
\beq \la{free-tau}
\phi\(x\) = \int {d^3 p \over \(2\pi\)^3}{1 \over 2\omega_p}\(b\(\mathbf p\)e^{ip\cdot x}+d^\dagger\(\mathbf p\)e^{-ip\cdot x}\)
\eeq
Where $b$ and $d^\dagger$ are the particles' annihilation and creation operators, and $\omega_p^2={\bf p}^2+m^2$. Using the RRC coordinates one can rewrite the phase factors as
\beq
x\cdot p=\tau\(\rho\,\hat {\mathbf x}\cdot\mathbf{p}-\omega_p\sqrt{1+{\rho^2}}\)\ ,
\eeq
we see that at large $\tau$ (late times) the integral in (\ref{free-tau}) is dominated by a saddle point at $\mathbf{p}_*=m\rho\, \hat {\mathbf x}$. Hence, 
\beq
\lim_{\tau \to \infty}\phi\(x\)\quad\propto\quad \tau^{-3/2}\[b\(m\rho\,\hat {\mathbf x}\)e^{-i\tau m}+id^\dagger\(m\rho\,\hat {\mathbf x}\)e^{i\tau m}\]
\eeq
This is the analog of (\ref{Phot2}) for massive particles, where at large $\tau$ they come / go from a unique direction in the sky. 

Under gauge transformation, the charged scalar field transforms as $\phi\(x\)\ \to\  e^{i\lambda\(x\)}\phi\(x\)$, where for convenience we have set the electric charge to one, ($e=1$). Correspondingly, the annihilation operator transforms as
\beq \la{mass_part_first}
b\(\mathbf p\)\  \to\  
e^{i\tilde\varepsilon({|{\bf p}|/ m},z_p,\bar z_p)}\,b({\bf p})\ .
\eeq

Finally, at large $\tau$, the generator of the LGT acts as
\beq \la{commutation_Massive}
\[Q_\ve^+, b\(\mathbf p\)\] = -b\(\mathbf p\)\int {d^2z\over4\pi} {{\gamma_{z,\bar z}}\,\ve(z,\bar z)\,{m^2}\over \(\hat {\mathbf x}_{\(z\)} \cdot {\bf p}-\sqrt{m^2+{\bf p}^2}\)^2}\ .
\eeq
The same transformation rule holds for the spin 1/2 electron.

\subsection{The LGT Generator}
Next, we obtain an expression for the LGT generator. Generally, the generator of a gauge transformation $\lambda$ is the integral of the corresponding current over a fixed time slice
\beq
Q^{(\lambda)} =\int\limits_{\text{fixed }t}\!\! d^3x\, j^{(\lambda)}_0\ ,\qquad\text{where}\qquad j^{(\lambda)}_\mu=\d^\nu\(\mathcal{F}_{\nu\mu}\lambda\)
\eeq
is the Noether current of the general gauge transformation characterized by $\lambda$. We can integrate over the radial coordinate $r'=|\mathbf x|$ to arrive at
\beq
Q^{(\lambda)} = \lim_{r' \to \infty}\int d^2z\,r'^2 \,\gam\,  \mathcal{F}_{rt}\(r', t, z, \bar z\)\lambda
\eeq
where we assumed that the transformation is regular at the origin, namely that $\lim\limits_{r \to 0,\, t\,\text{fixed}} \(r^2 \lambda\) = 0$. Changing to null retarded coordinates, we can represent the charge at the $u\to-\infty$ edge of future null infinity, see figure \ref{Penrose_Null}. Specializing to LGT (\ref{trans-null}) then yields,
\beq
Q_\ve^+ = \int\limits_{\mathscr{I}^+_-} d^2z\,r^2 \,\gam\,\ve(z,\bar z)\,  \mathcal{F}^{-}_{ru}\(r,u, z, \bar z\)
\eeq
where
\beq\la{Fpm}
\mathcal{F}_{rt}^{(\mp)}\(z, \bar z\)\equiv\lim_{u\to\mp\infty} \lim_{r\to\infty}\mathcal{F}_{rt}\(r, u, z, \bar z\)\ .
\eeq
Particularly, for a constant $\varepsilon(z,\bar z)$, the generator $Q_\ve^+$ coincides with the electric charge (the generator of the standard global $U(1)$ symmetry of QED). Utilizing integration by parts with respect to $u$, we represent $Q^+_\varepsilon$ as a sum of an integral over future time-like infinity and of a total derivative over future null infinity
\beq \la{Q-first}
Q_\ve^+ =\lim_{r\to\infty}\int d^2z\,r^2\,\gam\,\ve\zbz\[\mathcal{F}_{ru}^{(+)} -\int\limits_{\mathscr{I}^+} du\, \d_u \mathcal F_{ru} \]
\eeq

Next, we use the bulk equation of motion, 
\beq
j^M_u=\d^\nu \mathcal{F}_{\nu u}={1\over r^2\gam } \(\d_z\mathcal F_{zu} + \d_{\bar z}\mathcal F_{\bar z u}\)-\d_u \mathcal{F}_{ru}=-{\d_u\over r^2\gam } \(\d_z \mathcal A_{\bar z} + \d_{\bar z} \mathcal A_{z}\)-\d_u \mathcal{F}_{ru}
\eeq
to simplify the second term of (\ref{Q-first}),
\beq \la{gen-LGT}
Q_\ve^+ =\int d^2z\,\ve\zbz\[\gam r^2 \mathcal{F}_{ru}^{(+)}+ \int du\, \d_u \(\d_z \mathcal A_{\bar z}^{(\infty)} + \d_{\bar z} \mathcal A_{z}^{(\infty)} \) \]\ ,
\eeq
here,  we used the fact that massive particles cannot reach null infinity and (hence $j^M_u=0$ there). 
The first term in (\ref{gen-LGT}) is referred to as the ``hard'' part of the charge and the second is referred to as the ``soft'' part.

\subsection{Physical Asymptotic States}
The construction of physical asymptotic states in QED has been put forward by Kulish and Faddeev \cite{kulish}, following preceding papers by Chung \cite{Chung} and Kibble \cite{Kibble}. More recent outlook on the subject can be found in \cite{modern_states}. In this construction, the incoming and outgoing states are dressed by a coherent cloud of photons, so that they diagonalize the interacting asymptotic Hamiltonian.\footnote{This Hamiltonian includes, in particular, the slow decaying part of the Coulomb potential. As a result, unlike the bare S-matrix, the S-matrix between dressed states reduces through the semi-classical limit smoothly.}

The eigenstates obtained by this method factorize to the bare ``undressed'' particle and a soft ``cloud'' of photons
\beq \la{asym-particle}
|\Psi_{as}\>\equiv e^{-R_f}\hat \Psi|0\>\ ,
\eeq
where $\hat \Psi$ is an undressed creation operator from the naive Fock space. The soft photon dressing factor $e^{-R_f}$ added at a late time $t_0$ is given by  
\beq
R_f \equiv \int {d^3\mathbf p \,\hat \rho\(\mathbf p\)\over \(2\pi\)^3 2\omega_p}\int{d^3\mathbf k \over \(2\pi\)^32\omega_k}\[f\(\mathbf k, \mathbf p\)\cdot\e^{\alpha}(k)\, a_\alpha^\dagger\(\mathbf k\) - f^\star\(\mathbf k, \mathbf p\)\cdot \e^{\(-\alpha\)}(k)\, a_\alpha\(\mathbf k\)\]\ .
\eeq
here, a repeated index $\alpha$ means a summation over the two polarizations $\alpha =+$ and $\alpha=-$. Additionally,\footnote{If there is more than one charged matter field the creation and annihilation operators should be indexed appropriately.}
\beq
\hat \rho\(\mathbf p\) \equiv b^\dagger\(\mathbf p\)b\(\mathbf p\)-d^\dagger\(\mathbf p\)d\(\mathbf p\)
\eeq
is the charged matter density operator. {The function $f_\mu\(\mathbf k, \mathbf p\)$ takes the following form \cite{kulish}
\beq \la{form-function}
f_\mu\(\mathbf k, \mathbf p\) \equiv \({p_\mu \over k\cdot p}-{c_\mu \over \omega_k}\)\exp\(i{k\cdot p \over \omega_p}t_0\)\ ,% ,\ \text{where}\quad c_0 = {1\over 2\omega_k}\ ,\, c_i = {-k_i\over 2\omega^2_k}\ ,
\eeq
where $p$ is the charged matter on-shell four-momenta and $k$ is the photon on-shell (null) four-momenta}. A dot between two four-vectors is a shorthand for their product ($a\cdot b=a_\mu b^\mu$). {Finally, the four vector $c=c(k)$ is a $k$-dependent null four vector satisfying $c\cdot k=\omega_k$. A convenient parametrization of $c$ is,
\beq
c_\mu={\omega_k\,q(k)_\mu\over k\cdot q(k)}\ ,
\eeq
where $q\(k\)$ is a $k$ dependant projective null four vector. It is characterized by a map from the sphere to itself $w(z,\bar z)$, as $q=q_{w\bar w}$.}

One may commute $e^{R_f}$ through the annihilation operators, so that there remains a separate dressing for each electron,
\beq\la{qparametrization}
\<0|b\(\mathbf p\)e^{R_f}=\<0|e^{R_f(p)}b\(\mathbf p\)\ .
\eeq
Using the commutation relation $\[b\(\mathbf p\),\hat \rho\(\mathbf p'\)\]=\(2\pi\)^32\omega_p\delta^3(\mathbf p-\mathbf p')\,b\(\mathbf p\)$, we have,
\beq\la{Rfp}
R_f(p)= \int{d^3\mathbf k \over \(2\pi\)^32\omega_k} \[f\(\mathbf k, \mathbf p\)\cdot\e^{\alpha}(k)\, a_\alpha^\dagger\(\mathbf k\) - f^\star\(\mathbf k, \mathbf p\)\cdot \e^{\(-\alpha\)}(k)\, a_\alpha\(\mathbf k\)\]\ .
\eeq
The new operator acts only on the photons Fock space. Working with the $R_f(p_i)$'s instead of $R_f$ gives rise to an apparent freedom to assign different dressings for different particles. Namely, we denote the $c$-vector in the dressing of the $i$'th particle by $c_i$.

{In \cite{kulish}, all asymptotic states where dressed with the same $c$-vector.\footnote{The $c$-vector used in \cite{kulish} is $c_0 = -{1\over 2}$, $c_i = {k_i\over 2\omega_k}$. It corresponds to $q=(\omega_k,-\vec k)$ in (\ref{qparametrization}).} The scattering of such states with uniform $c$-vector is free of IR-divergences to all orders in perturbation theory.}

On the other hand, IR divergences do not generally cancel out in scattering of dressed states with different $c_i$-vectors. A simple generalization of \cite{Chung} shows that the interactions between photons in different dressings will yield non-canceling IR divergent factors unless the specific matrix element satisfies the condition
\beq\la{IRfreecond}
\sum_{i\in \text{incoming}}q_i\,c_i(\hat k)=\sum_{i\in \text{outgoing}}q_i\,c_i(\hat k)
\eeq 

When all the $c_i$'s are equal, (\ref{IRfreecond}) is trivially satisfied due to electric charge conservation. However, one can easily construct different $c_i$'s that satisfy (\ref{IRfreecond}). 

In the following sections we will see that (\ref{IRfreecond}) is equivalent to the conservation of the LGT charges. Amplitudes between states with different LGT charges, namely that violate (\ref{IRfreecond}), are expected to vanish due to exponentiation of the IR divergences.

Before we may continue and investigate the LGT properties of dressed states, we must address a small caveat in the way LGT in (\ref{gen-LGT}) are related to soft photons that is relevant to our discussion. When analyzing the soft behavior of an amplitude, one should consider the limit in which the energy of a photon approach zero $\omega_k\to0$. While a photon at strictly zero energy and momentum is nothing, the zero energy limit of an amplitude (which is a distribution) may be non-trivial. In accordance, when discussing LGT, one should implement an IR cutoff and take the soft limit only after calculations are done (and the IR regulator is removed). For example, one may employ a radial cutoff $r<r_{IR}$ and, at the end of the day, take $r_{IR}\to\infty$. However, here (and everywhere else as far as we know) we did not introduce such an IR cutoff. In particular, $A^{(\infty)}$ in (\ref{gen-LGT}) is evaluated at $r\to\infty$. Doing so amounts to treating photons of strictly zero energy as if they were physical particles. This introduces a known annoying factor of two. It comes about because at zero energy there is no way to distinguish between photons of opposite helicities, which engenders a double counting of these states. To fix this problem we introduce a step function
\beq
\Theta\(\omega_k\) \equiv
\left\{
\begin{array}{lr}
	1  &  \quad\omega_k \not \in \delta^\star\(0\) \\
	1/2 &  \quad\omega_k \in \delta^\star\(0\)
\end{array}
\right.
\eeq	
where $\delta^\star\(0\)$ is a neighborhood of $0$ much smaller than $1/t_0$, and rewrite the Kulish-Faddeev dressing factor as if strictly zero energy photons were physical states
\beq
R_f = \int {d^3\mathbf p \,\hat \rho\(\mathbf p\)\over \(2\pi\)^3 2\omega_p}\int{d^3\mathbf k \over \(2\pi\)^32\omega_k} \Theta\(\omega_k\)\[f\(\mathbf k, \mathbf p\)\cdot\e^{\alpha}(k)\, a_\alpha^\dagger\(\mathbf k\) - f^\star\(\mathbf k, \mathbf p\)\cdot \e^{\(-\alpha\)}(k)\, a_\alpha\(\mathbf k\)\]\ .\la{cloud}
\eeq
We leave a more proper treatment, utilizing an IR regulator, to the future.

In the following sections, we will incorporate all the ingredients reviewed above to compute the LGT of the physical vacuum and of the physical dressed asymptotic states. In accordance with the above, we will see that the LGT charge may be thought of as a measure of the vacua on top of which we build our amplitude.

\section{LGT of the Vacuum} \la{sec_lgtvac}

{When discussing LGT, the relaxed condition (\ref{fall-offs-a}) allows one to choose a vacuum by determining the zero energy states that annihilate it. In this subsection we show that acting with the generator of any LGT on the vacuum extended in the smooth fashion,
\beq\la{vacuumstate}
\[\lim_{\omega \to 0}a\(\omega\cdot \hat x\)\]|0\> = 0\ ,
\eeq
in the future or past gives a state that is orthogonal to any state constructed on the original vacuum.} Namely, we show that
\beq \la{Q-On-Vac}
\<0|Q_\ve^+\,\hat\Psi^{out}e^{R_f}\,\mathcal{S}\,e^{-R_f}\hat\Psi^{in}|0\> = 0\ .
\eeq
We recall that the LGT generator in (\ref{gen-LGT}) is made out of two pieces. The so called soft piece at future null infinity, and the hard piece at future infinity. When acting with the latter on a massive charged particle, it measures the phase it acquires under the LGT. However, when acting on the vacuum the hard part simply vanishes. The reason behind this is that photons can only go to (come from) null infinity and therefore cannot be annihilated (created) at future infinity. Consequently, when acting on the vacuum, we need to consider only the soft part of the LGT generator.

In order to avoid a more cumbersome presentation, we assume that $\varepsilon\zbz$ does not have a pole at $z=\infty$. Including such a pole does not lead to any complications and does not effect the result.

Using integration by parts with respect to $z$ and $\bar z$, the action of the generator (the soft part) on the vacuum gives
\beq
\<0|Q_\ve^+=-\int d^2z \int du\,\<0|\[\d_z\ve\(z,\bar z\) \d_u \Ascr_{u,\bar z}\(z,\bar z\)+\d_{\bar z}\ve\(z,\bar z\) \d_u \Ascr_{u,z}\(z,\bar z\)\]\ .
\eeq
As the photon is massless, the gauge field at some direction of future null infinity can be decomposed in creation and annihilation operators in that direction only. More explicitly, we derive in appendix \ref{AsymFieldAppe}\footnote{Note the factor of $1/2$ arising from the convention that under an integration over ${\mathbb R}_+$ we have
\beq\la{soft-delta}
\int\limits_0^\infty d\omega \delta\(\omega\)f\(\omega\)\equiv {1\over 2}f\(0\)\ .
\eeq}
\beq \la{softPhotons}
\int du\, \d_u \Ascr_{\bar z}\(u,z,\bar z\) = -\lim_{\omega \to 0}{\omega\sqrt{\gam} \over 8\pi}\[a_-\(\omega \hat {\mathbf x}_z\)+a_+^{\dagger}\(\omega \hat {\mathbf x}_z\)\]\ ,
\eeq
The subscript ($\pm$) stands for the two polarizations (\ref{polarizations}). A similar expression holds for $\d_u \Ascr_{z}$ (with the two polarizations interchanged). The limit $\omega\to0$ indicates that these are soft photons, hence the name ``soft'' part of $Q_\ve^+$. Since the creation operators annihilate the Bra vacuum, we remain with
\beq\la{vac_with_soft}
\<0|Q_\ve^+=\lim_{\omega \to 0}{\omega \over 8\pi}\int d^2z \sqrt{\gam}\,\<0| \[\d_{z}\ve\(z,\bar z\)a_{-}\(\omega \hat {\mathbf x}_z\)+\d_{\bar z}\ve\(z,\bar z\)a_{+}\(\omega \hat {\mathbf x}_z\)\]\ .
\eeq

Next, we demonstrate that scattering amplitudes between physical states with external soft photons are finite. Ergo, due to the extra overall factor of $\omega$ in (\ref{vac_with_soft}), {matrix elements {from a state constructed on $|0\>$} to the state $\<0|Q_\ve^+$ vanish in a physical scattering process.}

Consider a scattering amplitude between physical asymptotic states with one external soft photon 
\beq
\mathcal{M}_{q\text{ soft}}\equiv \lim_{\omega_q \to 0}\<0|a_-\(\mathbf q\)\hat \Psi^{out}e^{R_f}\ \mathcal S\ e^{-R_f}\hat\Psi^{in}|0\>
\eeq
where $\mathcal S=\lim\limits_{T\to\infty}e^{iT\,{\cal H}}$ is the S-matrix and $R_f$ is given in eq. \ref{cloud}. 
Here, $\hat \Psi^{in/out}$ are arbitrary undressed states that may include photons with energy $E>\omega_q$ only, and we used the fact that $R_f$ is anti-hermitian. Figure \ref{sPdiag} shows all the ways to connect the external soft photon to the amplitude. These are
\begin{figure}[t]
\centering
\def\svgwidth{5cm}
\includegraphics[width=.98\textwidth]{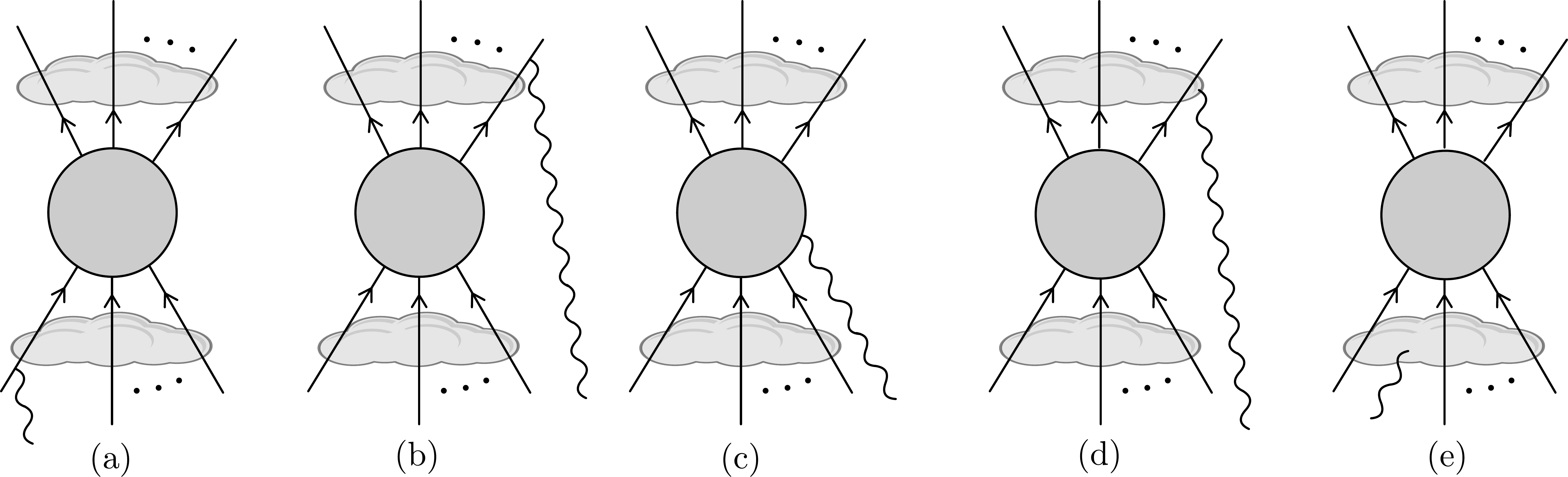}
\caption{All the ways to connect an external soft photon to a Feynman diagram. In (a)-(c) the soft photon connects to an external or internal leg. These are the traditional soft contributions. In (d) and (e) the soft photon is connected to the ``clouds'' dressing the asymptotic states.}\la{sPdiag}
\end{figure}
\begin{enumerate}
\item A soft photon connecting to an external incoming or outgoing leg, see figure \ref{sPdiag}.a and \ref{sPdiag}.b. As a result of the soft theorem, these give rise to a $\pm e{p^\mu_i \e_\mu^{-} \over p^\nu_i q_\nu}$ factor correspondingly. 
\item A soft photon connecting to an internal propagator, figure \ref{sPdiag}.c. This yields a finite result because it has no pole at the soft limit.
\item A soft photon connecting to the incoming or outgoing soft photon ``cloud'' $e^{-R_f}$, see figure \rf{sPdiag}.d and \ref{sPdiag}.e. Since any photon can be annihilated by an anti-photon of the same momentum in the cloud, these yield a {$\mp\sum_i e\({p_i\cdot \e^{-} \over p_i\cdot q}-{c_i\(\mathbf q\)\cdot \e^-\over \omega_q}\)e^{-i{qp_i \over \omega_{p_i}}t_0}$} factor correspondingly. This is a contribution that emerges from the dressing and does not appear in the usual soft theorem.
\end{enumerate}
Summing over the diagrams and taking the soft limit results in 
\beqa\la{Qvacuum}
\lim_{\omega_q \to 0}\omega_q\,\mathcal{M}_{q\text{ soft}}&\!=\!&\lim_{\omega_q \to 0}\omega_q\Big[\,\sum_{j \in \text{out}} {p_j\cdot \e^{-} \over p_j\cdot q}\(1-e^{i{q\cdot p_j \over \omega_{p_j}}t_0}\) - \sum_{i\in \text{in}} {p_i\cdot \e^{-} \over p_i\cdot q}\(1-e^{i{q\cdot p_i \over \omega_{p_i}}t_0}\)\\
&&\qquad\qquad\qquad{+\sum_{i\in \text{outgoing}}q_i{c_i(\hat q)\cdot\e^-\over\omega_q}-\sum_{i\in \text{incoming}}q_i{c_i(\hat q)\cdot\e^-\over\omega_q}}
\Big]\mathcal{M}\ ,\nn
\eeqa
where
\beq
\mathcal{M} \equiv \<0|\hat \Psi^{out}e^{R_f}\ \mathcal S\ e^{-R_f}\hat \Psi^{in}|0\>
\eeq
is the amplitude without the soft photon and $q_i$ is the charge of the i'th particle. Consider first the second line in (\ref{Qvacuum}). Since it is proportional to the condition (\ref{IRfreecond}) it contributes zero to any finite amplitude. Next, we consider the contributions from connecting the soft photon to the external legs and to the clouds. As $\omega_r\to0$, these two cancel each other causing this expression to vanish. This cancellation is one of the novelties of the physical asymptotic states of \cite{kulish}. We conclude that as a result of the soft theorem
\beq
\lim_{\omega_q \to 0}\omega_q\,\mathcal{M}_{q\text{ soft}} = 0\ .
\eeq
Returning to the expression for $\<0|Q_\ve^+$ in (\ref{vac_with_soft}) we see that
\beq \la{decouple}
\<0|Q_\ve^+\,\hat\Psi^{out}e^{R_f}\,\mathcal{S}\,e^{-R_f}\hat\Psi^{in}|0\> = 0
\eeq
for any two physical asymptotic states $e^{-R_f}\hat\Psi|0\>$.\footnote{In the first version of this paper we have falsely concluded from (\ref{decouple}) that the state $\<0|Q_\ve^+$ is null. Generically, The operator $Q_\ve^+$ generates a transformation between different super-selection sectors of the theory and hence, the state $\<0|Q_\ve^+$ is non-trivial. We thank Thomas Dumitrescu and Burkhard Schwab for pointing this out to us.\la{corr}}

In conclusion, we see that the S-matrix between the vacuum $|0\>$ in (\ref{vacuumstate}) and the state $\<0|Q_\ve^+$ vanishes. When working with undressed states is was established that the Ward identity of the LGT symmetry yielded the Soft theorem \cite{Y-M,Y-M-old,grav_1,grav_2,mass-lessQED,massQED,othermassQED}. Here we learn that it's merely an outcome of working with the wrong asymptotic states.\footnote{The soft theorem and not the soft expansion is the one yielded by the LGT Ward identity, meaning that the zero energy limit of the soft photons is taken first. See \cite{Cachazo:2014dia} for a discussion of the differences.}

{Note that in the calculation made in this section we assume that the singular term in the soft theorem does not change under renormalization, see \cite{Cachazo:2014dia} for discussion. We will not attempt to address the question of whether this assumption is correct here. However, as we have seen, the soft theorem and the Kulish-Faddeev dressing have the same form due to their common origin. Hence, we expect that if one changes under renormalization, the other will change as well, and in the same way.} 

\section{LGT of Dressed Particles} \la{sec_dressed}
We are now able to compute the LGT of a physical asymptotic state. We focus on a single particle state carrying momentum $p$ and dressed with the photon cloud (\ref{Rfp}). The action of the LGT generator (\ref{gen-LGT}) on the dressed asymptotic state (\ref{qparametrization}) decomposes to three pieces  
\beq\la{threepieces}
\<0|e^{-R_f(p)}b\(\mathbf p\)Q_\ve^+=\<0|Q_\ve^+\,e^{-R_f(p)}b\(\mathbf p\)+\<0|e^{-R_f(p)}[b\(\mathbf p\),Q_\ve^+]+\<0|[e^{-R_f(p)},Q_\ve^+]b\(\mathbf p\)\ .
\eeq
The second term was calculated in subsection \ref{massive-trans}. For convenience we repeat the result (\ref{commutation_Massive}) here
\beq
\[b\(\mathbf p\),Q_\ve^+\] = b\(\mathbf p\)\int {d^2z\over4\pi} {{\gamma_{z,\bar z}}\,\ve(z,\bar z)\,{m^2}\over \(\hat x_{\(z\)} \cdot {\bf p}-\sqrt{m^2+{\bf p}^2}\)^2}\ .
\eeq
The third and last term in (\ref{threepieces}) is the subject of this section. That is to say, we will now compute the action of LGT on the Kulish-Faddeev dressing factor 
\beq
\[R_f(p), Q_\ve^+\]=\!\int\!{d^3\mathbf k\, \Theta\(\omega_k\)\over 16\pi^3\omega_k} \(f\(\mathbf k, \mathbf p\)\cdot\e^{\alpha}(k)\[a_\alpha^\dagger\(\mathbf k\),Q_\ve^+\]- f^\star\(\mathbf k, \mathbf p\)\cdot \e^{\(-\alpha\)}(k)\[a_\alpha\(\mathbf k\),Q_\ve^+\]\)\ .\nonumber
\eeq
Using (\ref{commutation_photon_0}) for the action of $Q_\ve^+$ on the photon we arrive at
\beq\la{QRf}
\[R_f(p),Q_\ve^+\]= -\int{\sqrt{\gamma_{z\bar z}}\,d^2z\over 4\pi}\({p^\mu \over p\cdot\hat k_{z,\bar z}}-c^\mu\)\[\e^-_\mu(\hat k_{z,\bar z})\d_z \ve\(z,\bar z\)+\e^+_\mu(\hat k_{z,\bar z})\d_{\bar z} \ve\(z,\bar z\)\]\ ,
\eeq
where, according to our convention (see eq. (\ref{soft-delta})), the delta function yielded half the value of the integrand at $0$. Here, $\hat k_{z,\bar z}\equiv(1,\hat{\bf k}_{z,\bar z})$ is the null vector whose spatial component is a unit vector in the direction $(z,\bar z)$ and $\e^\pm$ are the photon polarization vectors.\footnote{The apparent correlation between the polarizations and the derivatives with respect to $z$ and $\bar z$ originate from the choice of these vectors in (\ref{polarizations}).} Next, we integrate by parts with respect to $z$ and $\bar z$ correspondingly. After some algebra one finds that the two terms in (\ref{QRf}) give the same result and hence, an overall factor of two
\beq\la{RQepsilon}
\[R_f(p),Q_\ve^+\]=\int{d^2z\over 4\pi}\(\gamma_{z\bar z}\,{p^2 \over (p\cdot\hat k_{z,\bar z})^2}-C\zbz\)\,\ve\(z,\bar z\)\ ,
\eeq
where,
\beq \la{Form-integral}
C\zbz = \,\[\d_z (c(\hat k)\cdot \e^-)+\d_{\bar z} (c(\hat k)\cdot \e^+)\]\ .
\eeq
Finally, we note that
\beq
{p^2 \over (p\cdot\hat k_{z,\bar z})^2}= {-m^2\over \( {\bf p} \cdot\hat{\bf k}_{z,\bar z}-\sqrt{m^2+{\bf p}^2}\)^2}\ ,
\eeq
which is proportional to the Lienard-Wiechert Field sensed approaching time-like future infinity in the direction $\hat k_{z,\bar z}$.

In conclusion, we see that the LGT of the bare particle and the $p$-dependent part of the LGT of the dressing cloud exactly cancel each other. {Hence, for the vacuum $\<0|$ defined in (\ref{vacuumstate}), we may measure the LGT charge flowing between any two physical asymptotic states,
\beq
{\<0|\hat \Psi_{as}^{out}Q_\ve S\,\hat \Psi_{as}^{in}|0\> \over \<0|\hat \Psi_{as}^{out}S\hat \Psi_{as}^{in}|0\>}=-\sum_{i\in\text{outgoing}}q_i\int {d^2z\over 4\pi} C_i\zbz \ve\zbz\ .
\eeq
We see that the value of the charge only depends on the choice of $c_i$ in the dressing (\ref{form-function}) through the relation (\ref{Form-integral}). As a result, the preservation of this charge is exactly condition (\ref{IRfreecond}). One may wonder if a more general conservation law exists. To answer that question, we retrace the steps of \cite{mass-lessQED,massQED,othermassQED,grav_1,grav_2,Y-M,Y-M-old} and seek a Ward identity for the LGT,
\beqa\la{wardusual}
0=\!\<0|\hat \Psi_{as}^{out}\,[Q_\ve,S]\,\hat \Psi_{as}^{in}|0\>\! \!\!\!&=&\!\!\! \lim_{\omega \to 0}\omega\(\<0|a_-\(\omega\hat x\)\hat \Psi_{as}^{out} \,S\, \hat \Psi_{as}^{in}|0\>+\<0|\hat \Psi_{as}^{out} \,S\, \hat \Psi_{as}^{in}\,a^\dagger_+\(\omega\hat x\)|0\>\)\nn\\
\!\!\!&=&\!\!\!2\lim_{\omega \to 0}\omega\,\<0|a_-\(\omega\hat x\)\hat \Psi_{as}^{out} \,S\, \hat \Psi_{as}^{in}|0\>\ ,
\eeqa
where we applied condition (\ref{IRfreecond}) and the preservation of electric charge. Also, in the last step, we used the soft photon crossing symmetry. Expectedly, this reproduces the result of the last section (\ref{decouple}). Another perspective is achieved by considering the vacua which are LGT eigenstates,
\beq\la{OmegaLambda}
\<\Omega_\Lambda| Q^+_\ve \equiv \int {d^2z \over 4\pi} \Lambda\zbz \ve\zbz \<\Omega_\Lambda|\ .
\eeq
They are related to the standard vacuum $\<0|$ (\ref{vacuumstate}) by
\beq \la{Q-modes}
\<0| = \int \mathcal D\[\Lambda\] e^{-{1 \over 2}\Lambda^2} \<\Omega_\Lambda|\ ,\qquad\text{where}\qquad\Lambda^2=\int {d^2z \over 4\pi}\,\Lambda^2(z,\bar z)\ .
\eeq
Now, physical asymptotic particles on top of the vacua \ref{Q-modes} are eigenstates of the LGT charge,
\beq\la{LGTtotal}
\<\Omega_\Lambda|e^{R_f(p)}b\(\mathbf p\)Q_\ve^+=\<\Omega_\Lambda|e^{R_f(p)}b\(\mathbf p\) {1 \over 4\pi}\int d^2z \(\Lambda\zbz-C\zbz\) \ve\zbz\ ,
\eeq
and the charge-scattering matrix commutation relation expectation value can be found explicitly,	
\beq \la{ward}
0=\!\<\Omega_\Lambda|\hat \Psi_{as}^{out}\,[Q_\ve,S]\,\hat \Psi_{as}^{in}|\Omega_{\tilde \Lambda}\>\! = \!\int {d^2z \over 4\pi} \(\Lambda\zbz - \tilde\Lambda\zbz\) \ve\zbz\<\Omega_\Lambda|\hat \Psi_{as}^{out} \,S\, \hat \Psi_{as}^{in}|\Omega_{\tilde \Lambda}\>\ .
\eeq
We have used the conservation of electric charge and condition (\ref{IRfreecond}) to cancel the $C$ factor in (\ref{LGTtotal}).} Since this is true for any $\ve\zbz$, the amplitude may be non-vanishing only if
\beq \la{result}
\tilde\Lambda\zbz = \Lambda\zbz\ .
\eeq
Note that since the amplitude $\<\Omega_\Lambda|\hat \Psi_{as}^{out}\,S\,\hat \Psi_{as}^{in}|\Omega_\Lambda\>$ is independent of the choice of $\Lambda$, the last equation (\ref{result}) simply means that the scattering starts and ends with the same vacuum.

Equations (\ref{ward}), (\ref{wardusual}) and (\ref{decouple}) are the main results of this paper. They mean that LGT charge that may be associated to the vacuum is conserved and independent of the particles momenta. In the next section we use Wilson lines to explain why the dressing has this effect.

\subsection{Wilson Lines Interpretation}\la{WL-int}

In this section, we show that the soft part of the dressing operator $R_f$ can be interpreted as a continuous set of Wilson lines. These Wilson lines have the effect of transporting the angular dependence of the LGT of any particle going to any direction in the sky, to a pattern of points on the sphere, that is independent of the particle's quantum numbers. This pattern is given by $C_i(z,\bar z)$ in (\ref{Form-integral}). In other words, it is determined by the choice of $c_i$ in the dressing of the particle (\ref{form-function}).\footnote{As noted before, since the $c_i$ part of the dressing commutes with the matter content and the rest of the dressing, any asymptotic state with any choice of $c_i$'s can be represented as a state for which all of the $c_i$'s of all of the particles are the same. The latter should be scattered on a vacuum rotated by their difference.}

We start with a spectral decomposition of the dressing operator
\beqa
R_f(p) \equiv \int d\omega_k\,{\mathbb R}_f(\omega_k,p)
\eeqa
As we have seen in (\ref{commutation_photon_0}), only the soft part of $R_f$ is relevant for the discussion of LGT. Hence, we will focus on it
\beq
\mathbb R_f(0,p)=\!\left. \int\!{d^2z\,\omega_k\gamma_{z\bar z} \over 16\pi^3}\Theta\(\omega\)\[f\({\mathbf k}_{z,\bar z}, \mathbf p\)\cdot\e^{\alpha}(k_{z,\bar z})\, a_\alpha^\dagger\(\mathbf k_{z,\bar z}\) - f^\star\({\mathbf k}_{z,\bar z}, \mathbf p\)\cdot \e^{\(-\alpha\)}(k_{z,\bar z})\, a_\alpha\({\mathbf k}_{z,\bar z}\)\]\right|_{\omega_k=0}.\nn
\eeq
Plugging in the polarization vectors defined in (\ref{polarizations}) and using the expression for the annihilation and creation operators in terms of the gauge field integrated at future null infinity (see (\ref{Phot-main})), we find the simple expression
\beq\la{simpleex}
\mathbb R_f(0,p)= -i\int {\sqrt{\gamma_{z\bar z}}\,d^2z\over 4\pi^2}\({p^\mu \over p\cdot\hat k_{z,\bar z}}-c^\mu\)\int du\(\e^-_\mu\Ascr_z+\e^+_\mu\Ascr_{\bar z}\)\ .
\eeq
Next, we use the relation ${1 \over 2\pi}\d_{\bar w}{1\over z-w}=\delta^2(w-z)$ and integration by parts to rewrite (\ref{simpleex}) as 
\beq
\mathbb R_f(0,p)={i \over 8\pi^3} \int d^2w\({\gamma_{w\bar w}\,p^2 \over (p\cdot\hat k_{w,\bar w})^2}-C\(w,\bar w\)\)\int du\int{d^2z} \({\Ascr_z \over \bar z-\bar w}+{\Ascr_{\bar z} \over z-w}\)\la{W-l-eq}\ .
\eeq
{In interpreting this expression, one should remember also the $1/2$ factor that arises from the convention set in eq. (\ref{soft-delta})}.

Finally, we rewrite the inmost integral as a sum of Wilson lines connecting the point $(w,\bar w)$ to infinity
\beq
\int {d^2z\over 4\pi} \({\Ascr_z \over \bar z-\bar w}+{\Ascr_{\bar z} \over  z-w}\)=\int{d\theta\over2\pi}\int dr A_r^{(\infty)}(u,re^{i\theta}+w,re^{-i\theta}+\bar w)\ ,
\eeq
where we have changed variables to $re^{i\theta}=z-w$. 
\begin{figure}[t]
	\centering
	\def\svgwidth{5cm}
	\includegraphics[width=.45\textwidth]{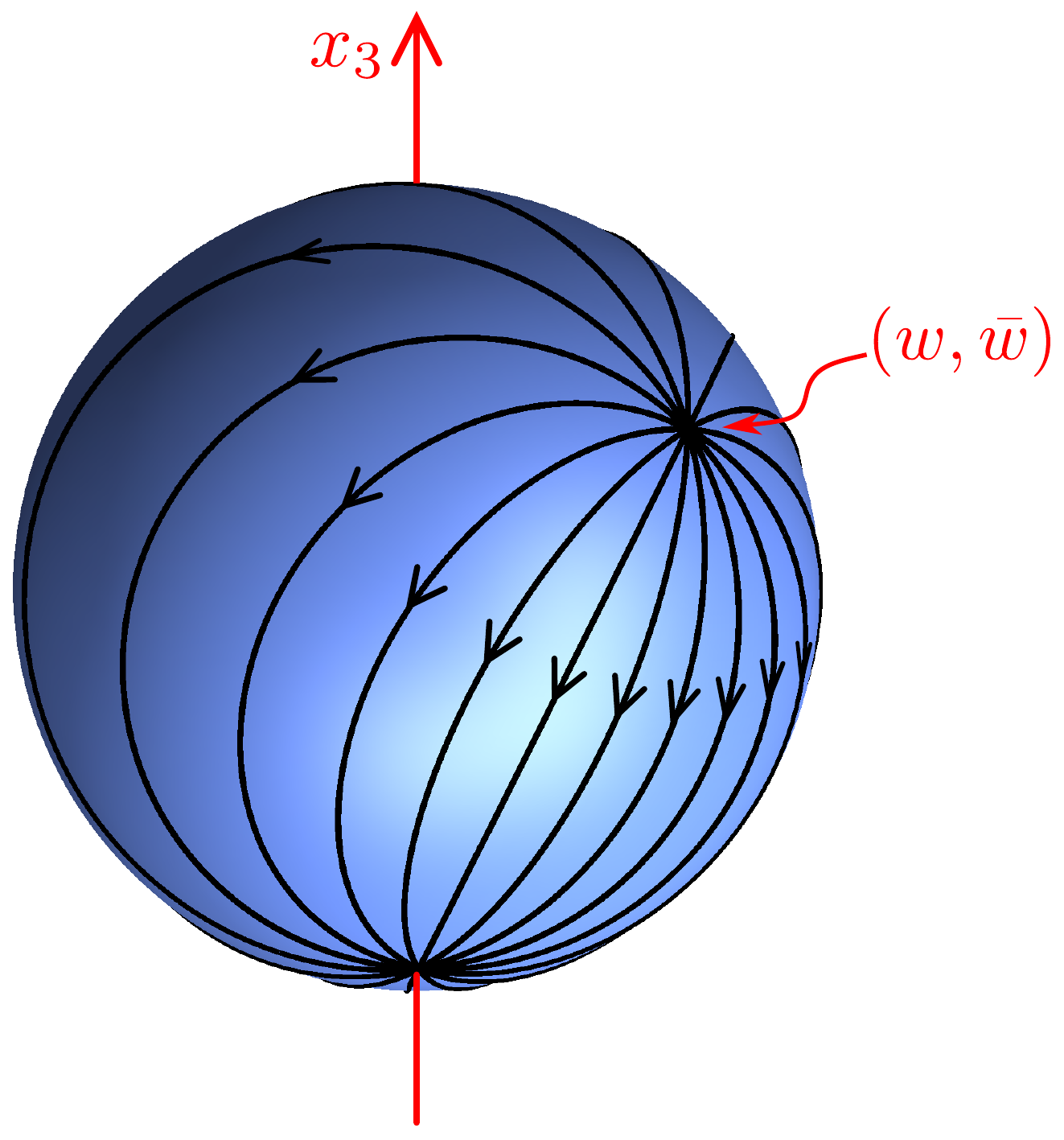}
	\caption{On the conformal sphere, $W_L\(w\)$ emerges from the point parametrized by $\(w,\bar w\)$ and reaches the south pole. For example, in this plot $w=e^{-i\pi/6}$.}\la{WLs}
\end{figure}

We see that for every angular direction $\theta$ around $w$ there is a radial Wilson line going to infinity as in figure \ref{WLs}. The Wilson lines emerging from each such point $w$ carry a fraction of the electron's charge to $w=\infty$. This fraction is nothing but the flux of the Lienard-Wiechert electric field created by a particle with momentum $\mathbf p$ through the opening angle $\gamma_{w\bar w}dwd\bar w$.
In addition, there are Wilson lines emerging from infinity in every angular direction $\theta$ towards every point on the sphere, carrying a fraction of the electric charge parametrized by (\ref{Form-integral}).
As a result, instead of having a complicated angular dependence, the LGT of any particle is independent of its momentum and is fully determined by the $C$-weighed average value of $\ve\zbz$ on the sphere
\beq \la{wehi-ave}
{1\over 4\pi}\int d^2z\, C\zbz \ve\zbz\ ,
\eeq
see (\ref{LGTtotal}).

To summarize, we observe that one of the roles of the Kulish-Fadeev dressing factor is, for any LGT, to ``standardize'' the dependence of all particles to one transformation rule. The total effect of a LGT factorizes to a standard global $U(1)$ transformation on asymptotic particles and a rotation of the vacuum.

\section*{Conclusions}
The global $U(1)$ charge in QED is the generator of gauge transformations that approach a non-zero constant at infinity. Large gauge transformations are generalization of the these global transformation. Instead of admitting a constant at infinity, their value  depends on the direction on the celestial sphere. In this paper we have explicitly shown that physical asymptotic charged particles transform under LGT in a way that does not depend on their momenta. In other words, they transform independently of their asymptotic direction. This is a result of a cancellation between the LGT of a bare charged asymptotic particle and (the non-global part of) the LGT of the Kulish-Fadeev soft photon cloud dressing it. Moreover, one may choose to assign the LGT charge of asymptotic particles to the vacuum, while the particles are all neutral.

Scattering on top of different vacua are related by symmetry and, hence, are the same. When working with undressed particles, the physical S-matrix implication of the spontaneous broken symmetry is the Weinberg Soft Theorem. Here we have shown that this is merely an outcome of working with non-physical asymptotic states. Once we consider the physical S-matrix, the implication of the spontaneous broken symmetry becomes trivial to leading order in the large volume limit. We leave the question of wether there are non trivial implications at sub-leading orders for future study.

That said, LGT still have an important role in QED. For example, they are strongly connected to Weinberg soft theorem. Also, they are a useful tool to parametrize memory \cite{suss,Pasterski:2015tva}.\footnote{In appendix \ref{memoryapp} we show how the memory effect manifests when working with physical asymptotic states.}

Although in this paper we only focused on QED, we believe our result generalizes to the BMS super-translation symmetry of gravity in {asymptotically} flat spacetime. Similar to the physical QED states considered in this paper, physical asymptotic states were constructed for gravity in \cite{grav_coherent}. {This very interesting extension to BMS super-translations would corroborate with \cite{other} and is left for future work.}

\section*{Acknowledgments}
We thank Freddy Cachazo, Massimo Porrati, Zohar Komargodski, Shimon Yankielowicz and Alexander Zhiboedov
for useful discussions. We especially thank Freddy Cachazo for pointing us towards
the work of Kulish-Fadeev on asymptotic states. {We thank Thomas Dumitrescu and Burkhard Schwab for pointing a wrong statement in the first version of this paper (see footnote \ref{corr}).} A.S. has been supported by the I-CORE
Program of the Planning and Budgeting Committee, The Israel Science Foundation (grant
No. 1937/12) and the EU-FP7 Marie Curie, CIG fellowship.

\appendix

\section{Asymptotic Field Expression}

\la{AsymFieldAppe}
In this appendix we derive a simple expressions of the photon creation and annihilation operators in terms of the gauge field at null infinity. Such simple relation only applies to massless fields because their plan waves are completely localized on the conformal sphere at null infinity.

We start with the fixed time expression for the canonically quantaized gauge field
\beq\la{free-field}
\mathcal A_\mu=\int{d^3k \over \(2\pi\)^3}{1 \over 2\omega_k}\(\e^{\alpha\star}_\mu\(\mathbf k\)a_\alpha^{\mathscr{I}^+}\(\mathbf k\)e^{ikx}+\e^{\alpha}_\mu\(\mathbf k\)a_\alpha^{\mathscr{I}^+}\(\mathbf k\)^{\dagger}e^{-ikx}\)\ ,
\eeq
where the sum over the polarization index $\alpha$ is over the values $\alpha=+,-$. Next, we decompose the plane waves in terms of spherical waves as
\beq\la{plane_wave_expansion2}
e^{ikx} = 4\pi e^{-i\omega\(u+r\)} \sum_{l=0}^{\infty}{i^l j_l(\omega r)}\sum_{m=-l}^{l}Y_l^m(\hat{\mathbf k})\,Y_l^{m*}(\hat{\mathbf x}) 
\eeq
and take the large $r$ limit at a fixed retarded time $u$. In this limit the Bessel function in (\ref{plane_wave_expansion2}) simplifies as
\beq\la{bessel_at_infty}
\lim_{x \to \infty} j_l(x) = -\frac{\sin(\frac{\pi l}{2}-x)}{x} + O(1/x^{2})\ .
\eeq
Using the completion relation of spherical harmonics
\beq\la{sphericalcomp}
\sum_{l=0}^{\infty}\sum_{m=-l}^{l}Y_l^m(\hat {\mathbf k})Y_l^{m\star}(\hat {\mathbf x})=\delta^2(\hat {\mathbf k}-\hat {\mathbf x})\ ,
\eeq
where 
\beq
\delta^2\(\hat {\mathbf x} - \hat {\mathbf k}\) \equiv \delta\(cos\theta_x - cos\theta_k\)\delta\(\phi_x-\phi_k\) =\delta^2\(z_x-z_k\)/\gamk\ .
\eeq
we see that
\beq\la{plane_wave_expansion}
\lim_{r \to \infty} e^{ikx} =  {2\pi i \over \omega r}\[e^{-i\(\omega-i\e\) \(u+2r\)}\delta^2(\hat{\mathbf {x}}+\hat{\mathbf {k}})-e^{-i\(\omega-i\e\) u}\delta^2(\hat{\mathbf {x}}-\hat{\mathbf {k}})\]\ . 
\eeq
\\
where we explicitly wrote the Feynman $i\e$ prescription to determine the dominant contribution. By plugging (\ref{plane_wave_expansion}) back into (\ref{free-field}) we learn that  
\beq
\begin{split}
\lim_{r \to \infty} \mathcal A_\mu\(r,u,\hat{\bf x}\) = -{i \over r}\int\limits_0^\infty&{d\omega \over 8\pi^2}\(\e^{\alpha\star}_\mu\(\hat {\mathbf x}\)a_\alpha\(\omega \hat {\mathbf x}\)e^{-i\omega u}-\e^{\alpha}_\mu\(\hat {\mathbf x}\)a_\alpha^\dagger\(\omega \hat {\mathbf x}\)e^{i\omega u}\)\ .
\end{split}
\eeq
In particular, for the $z$ and $\bar z$ components of the gauge field, $\mathcal A_z={\partial_z x^\mu}\mathcal A_\mu
$ we have
\beq
\lim_{r \to \infty} \mathcal A_z\(r, u, \hat x\)= -i\sqrt{\gamma_{z_{\hat{\bf x}}\bar z_{\hat{\bf x}}}}\int\limits_0^\infty{d\omega \over 8\pi^2}\(a_+\(\omega \hat {\mathbf x}\)e^{-i\omega u}-a_-^\dagger\(\omega \hat {\mathbf x}\)e^{i\omega u}\) 
\eeq
where $\hat {\mathbf x}_{z,\bar z}$ is a unit vector in the direction defined by the sphere coordinates $z$ and $\bar z$, see (\ref{inversetrans}). Inverting this relation, we arrive at
\beqa \la{Phot}
a_{+/-}\(\mathbf k\) &=& {4\pi i \over \sqrt{\gamk}}\int du\, e^{i\omega_k u} \Ascr_{z/\bar z}\(u, z_k, \bar z_k\)\ ,\\
a_{-/+}^\dagger\(\mathbf k\) &=& {-4\pi i \over \sqrt{\gamk}}\int du\, e^{-i\omega_k u}  \Ascr_{z/\bar z}\(u, z_k, \bar z_k\)\ ,\la{Phot2}
\eeqa
where $\omega_k = +|{\bf k}|$.

\section{Memory Effect of Physical Asymptotic States}\la{memoryapp}

One of the important properties of physical asymptotic states is that the transition of their quantum physics to the classical regime is smooth. In this appendix we show how the memory effect manifests itself when working with these states. The reason we have chosen to present this here is that, already at the classical theory, the memory effect may lead one to think that the memory effect is contradictory to the results presented in these notes. As we explain below, this is not the case.\footnote{We thank A. Zhiboedov for raising this question.}

We consider the experiment described in \cite{suss}, for which the measured memory operator is {(see equation (1.4) in \cite{suss}))}
\beq\la{memory}
\widehat O\(z,\bar z\) \equiv \lim_{r' \to \infty} -{1 \over \gam r'^2} \int dt\, \partial_t\(\d_z A_{\bar z} + \d_{\bar z} A_z\)\ .
\eeq
Here, $r'$ is the Cartesian radial coordinate. It specifies an asymptotically large sphere, such that the massive particles never reach it. The experiment considered in \cite{suss} only involves the outgoing state. More generally, memory is an in-in (or out-out) Schwinger-Keldysh type experiment. Accordingly, here we shall only focus on the outgoing state, and  assume zero initial charge. Since only massless particles reach the sphere at large $r'$, we can measure the action of this operator on the outgoing states at future null infinity, where it takes the form
\beq \la{memOp}
\widehat O\(z,\bar z\) \equiv \lim_{r \to \infty} -{1 \over \gam r^2} \int du\, \partial_u\(\d_z A_{\bar z} + \d_{\bar z} A_z\)\ .
\eeq

Using Gauss' law, (\ref{memory}) can be written as (see equation (2.2) in \cite{suss}))
\beq
\widehat O\(z,\bar z\) = -\int dt \d_{r'} E_{r'}\(z,\bar z\)\ ,
\eeq
where the time intgral is over the time when the detector was present, and $E_{r'}$ is the radial component of the electric field. Changing to retarded coordinates we see that
\beq
\widehat O\(z,\bar z\) = -\int du \(\d_r-\d_u\)\(E_r\(z,\bar z\)-E_u\(z,\bar z\)\)=\Delta E_r\zbz
\eeq
here, we used the fact that both $E_u$ vanishes and $E_r \propto O\(r^{-2}\)$ at $\mathscr I^+$. As mentioned earlier, in this experiment contributions to $\Delta E_r$ are made by the out-going particles only, and the initial charge is zero. Hence,
\beq
\widehat O\(z,\bar z\) = \sum_iE^+_r\(q_i;\mathbf p_i/m;z,\bar z\)\ ,
\eeq
where $E^+_r\(q;\mathbf p/m;z,\bar z\)$ is the radial component of the electric field at future null infinity that was emitted by a particle with momentum $\mathbf p$ and charge $q$.

Next, we will obtain a quantum expression for the memory effect, and show that it's equal to the classical one. Using equation (\ref{softPhotons}) and an integration by parts, 
the memory operator can be expressed in terms of soft photon creation and annihilation operators
\beq\la{memoryop}
\widehat O\(z,\bar z\) = -\lim_{\omega \to 0} {\omega \over 8\pi} \int d^2w \[\d_{\bar z}\delta^2\(w-z\) \sqrt{\gamma_{w,\bar w}} \(a_+ \(\omega\hat x_{w,\bar w}\)+a_-^\dagger \(\omega\hat x_{w,\bar w}\)\)+h.c.\]\ .
\eeq

For physical asymptotic states the memory operator takes the form
\beq
\<\Psi| e^{R_f}\widehat O\(z,\bar z\)  =  \<\Psi|\[e^{R_f}, \widehat O\(z,\bar z\)\] + \<\Psi| \widehat O\(z,\bar z\) e^{R_f}\ .
\eeq
When working with such particle states, the interaction of a soft photon with the bare particles and with the dressing cloud cancel each other as shown in section \ref{sec_lgtvac}. Therefore, there is no soft pole, and due to the factor of $\omega$ in (\ref{memoryop}) we can set
\beq
\<\Psi| \widehat O_z\(z,\bar z\)e^{R_f} = 0\ .
\eeq
that means that the memory operator depends only on the contribution of the dressing
\beqa \la{softMemory2}
\[e^{R_f}, \widehat O\(z,\bar z\)\]\!\!&=&\!\! \sum_i{q_i \over 8\pi}\[\d_{z} \(\sqrt{\gamma_{z,\bar z}}\,f\(p_i,\hat x_{z,\bar z}\) \cdot \epsilon^-\) +\d_{\bar z} \(\sqrt{\gamma_{z,\bar z}}\,f\(p_i,\hat x_{z,\bar z}\) \cdot \epsilon^+\)\]e^{R_f}\nn\\
\!\!&=&\!\!\sum_i\[E^+_r\(q_i;\mathbf p_i/m;z,\bar z\)\]e^{R_f}\ .
\eeqa
here, we used the fact that for the suggested experiment the initial electric charge, and hence the final electric charge, is $0$. Thus, the measurement will be exactly the same as in the classical limit.

One may regard the dressing factor as re-adding the asymptotic electric field to the particle. As a result, the soft photon theorems becomes trivial,\footnote{Here one should make a distinction between a soft photon theorem and a soft expansion in the sense of \cite{Cachazo:2014dia}.} but the dressing (taking the role of the quantum electric field) yields the memory.

This memory operator, however, is not the aforementioned generator of LGT. In the massive case, the latter reads (\ref{gen-LGT})
\beq
Q_\ve^+ =\int d^2z\,\ve\zbz\[\gam\, r^2 \mathcal{F}_{ru}^{(+)}+ \int du\, \d_u \(\d_z \mathcal A_{\bar z}^{(\infty)} + \d_{\bar z} \mathcal A_{z}^{(\infty)} \) \]\ .
\eeq
While $Q_\ve^+$ is a conserved charge for any $\varepsilon$, it's action on the matter operators is trivial in the sense discussed in the main text, (it can be thought of as the constraint that picks out the dressed states).

In summary, the memory effect does not change when working with physical asymptotic states. Furthermore, the measured ``memory'' is only the ``soft'' part of the LGT generator $Q^+_\ve$ (in the notation of (\ref{gen-LGT})). When calculating the LGT charge, the soft part, which may be non-trivial, is canceled out by the ``hard'' part.

The memory measurement for states with nonzero total electric charge is less straightforward. Combined with measurement of the outgoing particles, it may supply us with a measurement of the underlying vacuum.

\end{document}

%% file: Penrose_null2.pdf_tex
%% Creator: Inkscape 0.91_64bit, www.inkscape.org
%% Accompanies image file 'Penrose_null2.pdf' (pdf, eps, ps)
\begingroup%
  \makeatletter%
  \providecommand\color[2][]{%
    \errmessage{(Inkscape) Color is used for the text in Inkscape, but the package 'color.sty' is not loaded}%
    \renewcommand\color[2][]{}%
  }%
  \providecommand\transparent[1]{%
    \errmessage{(Inkscape) Transparency is used (non-zero) for the text in Inkscape, but the package 'transparent.sty' is not loaded}%
    \renewcommand\transparent[1]{}%
  }%
  \providecommand\rotatebox[2]{#2}%
  \ifx\svgwidth\undefined%
    \setlength{\unitlength}{864.65830078bp}%
    \ifx\svgscale\undefined%
      \relax%
    \else%
      \setlength{\unitlength}{\unitlength * \real{\svgscale}}%
    \fi%
  \else%
    \setlength{\unitlength}{\svgwidth}%
  \fi%
  \global\let\svgwidth\undefined%
  \global\let\svgscale\undefined%
  \makeatother%
  \begin{picture}(1,0.61355975)%
    \put(0,0){\includegraphics[width=\unitlength,page=1]{Penrose_null2.pdf}}%
    \put(0.04120795,0.56376433){\color[rgb]{0,0,0}\makebox(0,0)[lb]{\smash{$i^+$}}}%
    \put(0.04120795,0.0197344){\color[rgb]{0,0,0}\makebox(0,0)[lb]{\smash{$i^-$}}}%
    \put(0.15278329,0.44734286){\color[rgb]{0,0,0}\makebox(0,0)[lb]{\smash{$\mathscr{I}^+$}}}%
    \put(0,0){\includegraphics[width=\unitlength,page=2]{Penrose_null2.pdf}}%
    \put(0.34097955,0.27324495){\color[rgb]{0,0,0}\makebox(0,0)[lb]{\smash{$r'$}}}%
    \put(0.2947185,0.3065529){\color[rgb]{0,0,0}\makebox(0,0)[lb]{\smash{$i^0$}}}%
    \put(0.0079,0.59892273){\color[rgb]{0,0,0}\makebox(0,0)[lb]{\smash{$t$}}}%
    \put(0.23006177,0.43828947){\color[rgb]{0,0,0}\makebox(0,0)[lb]{\smash{$\Blue{r}$}}}%
    \put(0,0){\includegraphics[width=\unitlength,page=3]{Penrose_null2.pdf}}%
    \put(-0.00069466,0.56539733){\color[rgb]{0,0,0}\makebox(0,0)[lb]{\smash{$\red{u}$}}}%
    \put(0,0){\includegraphics[width=\unitlength,page=4]{Penrose_null2.pdf}}%
    \put(0.16931715,0.00374729){\color[rgb]{0,0,0}\makebox(0,0)[lb]{\smash{$(a)$}}}%
    \put(0,0){\includegraphics[width=\unitlength,page=5]{Penrose_null2.pdf}}%
    \put(0.6868215,0.56357794){\color[rgb]{0,0,0}\makebox(0,0)[lb]{\smash{$i^+$}}}%
    \put(0.6868215,0.01954799){\color[rgb]{0,0,0}\makebox(0,0)[lb]{\smash{$i^-$}}}%
    \put(0.79839674,0.44715647){\color[rgb]{0,0,0}\makebox(0,0)[lb]{\smash{$\mathscr{I}^+$}}}%
    \put(0.98659304,0.27305856){\color[rgb]{0,0,0}\makebox(0,0)[lb]{\smash{$r'$}}}%
    \put(0.94033199,0.30636652){\color[rgb]{0,0,0}\makebox(0,0)[lb]{\smash{$i^0$}}}%
    \put(0.65351354,0.59873634){\color[rgb]{0,0,0}\makebox(0,0)[lb]{\smash{$t$}}}%
    \put(0.87567528,0.43810309){\color[rgb]{0,0,0}\makebox(0,0)[lb]{\smash{$\Blue{\rho}$}}}%
    \put(0.64491888,0.56521094){\color[rgb]{0,0,0}\makebox(0,0)[lb]{\smash{$\red{\tau}$}}}%
    \put(0,0){\includegraphics[width=\unitlength,page=6]{Penrose_null2.pdf}}%
    \put(0.81697186,0.00374729){\color[rgb]{0,0,0}\makebox(0,0)[lb]{\smash{$(b)$}}}%
    \put(0,0){\includegraphics[width=\unitlength,page=7]{Penrose_null2.pdf}}%
  \end{picture}%
\endgroup%